# Fast Statistical Timing Analysis for Circuits with Post-Silicon Tunable Clock Buffers

Bing Li, Ning Chen, Ulf Schlichtmann
Institute for Electronic Design Automation, Technische Universitaet Muenchen, Germany
{b.li, ning.chen, ulf.schlichtmann}@tum.de

*Abstract*—Post-Silicon Tunable (PST) clock buffers are widely used in high performance designs to counter process variations. By allowing delay compensation between consecutive register stages, PST buffers can effectively improve the yield of digital circuits. To date, the evaluation of manufacturing yield in the presence of PST buffers is only possible using Monte Carlo simulation. In this paper, we propose an alternative method based on graph transformations, which is much faster, more than 1000 times, and computes a parametric minimum clock period. It also identifies the gates which are most critical to the circuit performance, therefore enabling a fast analysis-optimization flow.

## I. Introduction

Process variations have become larger in recent technology nodes. This trend makes the traditional worst-case timing analysis too pessimistic, which causes expensive overdesign and deprives designers of the valuable performance-yield information. To model the timing characteristics of a circuit more accurately, Statistical Static Timing Analysis (SSTA) has gained much attention in the research community in past years. This method models process variations with random variables directly and computes the complete performance-yield curve of the circuit. According to the assumption of the distributions of process variations and the method to model gate delays as well as the computation of the circuit delay, statistical timing algorithms can roughly be classified into several groups. First-order methods [1]–[3] use the canonical linear form [3] to represent gate delays and arrival times so that the computation can be simplified, at the expense of accuracy. To improve modeling and propagation accuracy, quadratic methods are proposed in [4]–[7], using second-order polynomials to approximate gate delays and arrival times. Other methods, for example [8], [9], can handle non-Gaussian process variations and the corresponding timing propagation.

The research on statistical timing analysis focuses mainly on combinational circuits, which implies that it is only applicable to flip-flop based circuits. However, many methods have also been deployed in industry to counter process variations, thus creating circuits with special structures. Post-Silicon Tunable clock buffers, PST buffers henceforth, are widely used in high performance designs, for example [10], [11]. The tunable or programmable buffers are inserted in the clock network to the registers which are relevant to critical paths. After manufacturing, the delay values of these buffers are adjusted, through the TAP port, to assign the critical paths more timing margin by allowing the delay compensation between register stages, therefore revitalizing the chips which may have failed to meet the timing specification. Consequently, with PST buffers the yield of the circuit should be higher than without them deployed.

Several methods have already been proposed for statistical timing analysis and optimization of circuits with PST buffers. In [12] a clock scheduling method is developed and PST buffers are selectively inserted to balance the skew that could be caused by process variations. Further in [13] algorithms are proposed to minimize the area required for the insertion of PST buffers, or to minimize the number of PST buffers in the circuit. In these methods, the yield of the circuit with PST buffers is computed using Monte Carlo simulation, therefore requiring much runtime. In [14] the yield loss due to process variations and the total cost of PST buffers are formulated together for gate sizing. The resulting optimization problem is solved using a stochastic cutting plane method mixed with a Monte Carlo based STA scheme, still having a slow convergence. Additionally, the placement of PST buffers is investigated in [15] and large benefit is observed when the clock tree is designed using the proposed tuning system. To the best of our knowledge, all existing methods depend on Monte Carlo based yield evaluation, thus causing considerable large runtime.

The research on statistical timing analysis and optimization has shown the advantage of using PST buffers in high performance designs. However, two problems still remain. The first is the requirement of a fast statistical timing analysis method, with which the runtime of the already proposed methods for optimization can be reduced. These methods still use Monte Carlo simulation to compute the yield of the circuit and no faster solution exists. The second is to select a small set of gates for sizing after the main circuit structure is stable. The selected gates should have high probabilities to affect the circuit performance, which can be improved by sizing these gates. In statistical timing analysis this probability is usually called criticality, and many methods have been proposed to describe and compute the criticalities of the gates in the circuit efficiently. In [3] the concept of criticality is first explored but without considering correlation. In [16] the sensitivities of gate and path delays to the circuit delay are computed. In [17] the criticality is computed using a cutset based method with a binary tree partition. Furthermore in [18] a fast criticality computation method is proposed with incremental yield gradients. Additionally in [19] a clustering based pruning is proposed to speed up the computation and improve the accuracy of criticalities. For ranking criticalities the computation of the maximum of a set of random variables is investigated in [20]. These methods, though accurate and fast, do not consider PST buffers, which allow the compensation of path delays across register stages and therefore make the criticality computation more complicated.

In this paper, we propose a fast method to compute the circuit delay in the presence of PST buffers. We also investigate the criticalities of gate delays when timing compensation is allowed across register boundaries. The main contributions of this paper are as follows.

- The proposed method computes a parametric minimum clock period for the circuit with PST buffers. The statistical properties of this minimum clock period, such as mean and variance, are directly available so that the yield of the circuit for any given clock period can be evaluated very fast. Since the computed circuit performance is in a parametric form, it can easily be integrated into other optimization methods for circuits with PST buffers.
- The proposed method is much faster, more than 1000 times, than Monte Carlo simulation by handling the path delay compensation across registers with a graph transformation based loop evaluation.
- The criticalities of gate delays in such circuits are defined and

computed. The gates with large criticalities are candidates for sizing when the circuit structure is stable. The proposed method can capture the most critical gates within very short runtime, therefore enabling a fast analysis-sizing cycle.

The rest of this paper is organized as follows. In Section II we give an overview and formulate the problem of timing analysis for circuits with PST buffers. In Section III we explain our statistical timing analysis and criticality computation method. Finally we show experimental results in Section IV and conclude our work in Section V.

## II. TIMING ANALYSIS WITH PST BUFFERS

In this section, we describe the timing constraints of digital circuits with PST buffers. The proposed method handles circuits in which all registers are edge triggered Flip-Flops (FFs). For simplicity we will only discuss setup time constraints. Hold time constraints can be analyzed similarly as formulated in [21], or met by delay insertion or padding, for example in [22].

In a digital circuit the clock signal is routed to FFs through a clock distribution network. If PST buffers exist on the clock routing to FFs, the clock signal reaches FFs not at the same time. Fig. 1 shows an example of two FFs with PST buffers.

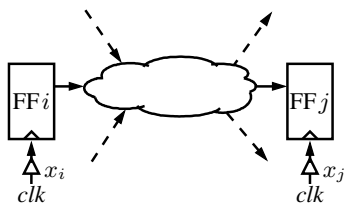

Fig. 1. FFs with PST Buffers

Assuming the common clock signal switches at reference time 0, the clock events at FF $i$ and $j$ happen at time $x_i$ and $x_j$ due to the PST buffers. To guarantee the setup time constraint of FF $j$, the minimum clock period $T$ must meet

$$x_i + d_{ij} \leq T + x_j - s_j \iff x_j - x_i \geq w_{ij} - T \quad (1)$$

for any pair of FFs $i$ and $j$, where $x_i$ and $x_j$ are the delay value of the PST buffers; $d_{ij}$ is the maximum delay of the combinational paths from $i$ to $j$; $T$ is the clock period; $s_j$ is the setup time of $j$; and $w_{ij} = d_{ij} + s_j$. Each PST buffer has a feasible delay range determined by design specification or the optimization algorithm, for example [13], written as

$$-r_i \leq x_i \leq r_i \quad (2)$$

where $r_i$ is a predetermined constant for the PST buffer to FF $i$. $x_i$ is defined in respect to a reference time so that it can be negative.

When process variations are considered, the values of $d_{ij}$ and $s_j$ can only be known after manufacturing. An individual chip is considered working if a set of values for $x_i$ and $x_j$ meeting (1) and (2) can be found. The PST buffers are adjusted to have delays of these values so that the given clock period $T$ can be met. This is actually a skew schedule problem in static timing analysis [23] to improve the circuit performance by cycle stealing, and can be solved by minimizing $T$ subject to (1) and (2), using linear programming or the binary search method in [21]. If the minimum clock period $T$ is smaller than the specification, the values of PST buffers, as $x_i$ and $x_j$ in Fig. 1, are determined accordingly, using linear programming, for example. Otherwise, the chip is considered failed. The goal of statistical timing analysis for circuits with PST buffers is thus to determine the percentage of chips for which a valid set of PST values can be found.

Since each chip after manufacturing can be considered as a sample in the parameter space, all minimum clock periods of the chips can be described using a random variable $T_{min}$. The yield of the circuit for a given clock period $T$ is computed by the probability $prob\{T_{min} \leq T\}$. Before manufacturing, $T_{min}$ can be computed by sampling the variables $d_{ij}$ and $s_j$ in (1) and computing the minimum clock period for each sample using Monte Carlo simulation. However this method needs a very long runtime for a large number of samples to guarantee simulation fidelity. In our method we directly compute $T_{min}$ from the constraints of (1) and (2) to avoid the runtime problem.

Because in each constraint of (1) and (2) there are no more than two variables $x_i$ or $x_j$, they together form a difference constraint problem. The constraint that there is a solution for the problem described by (1) and (2) is equivalent to the constraint that all the loops in the corresponding constraint graph are nonpositive [24, Ch. 25.5]. The constraint graph contains a node for each FF, corresponding to a variable $x_i$ or $x_j$ in (1) and (2). If a constraint (1) exists for FFs $i$ and $j$, a directed edge is created from $i$ to $j$, with weight $w_{ij} - T$. To incorporate the constraint (2), a common root node is created and an edge is created from it to $i$ with weight $-r_i$ for $x_i \geq -r_i$ and an edge from $i$ to the root node with weight $-r_i$ for $x_i \leq r_i$, equivalent to $-x_i \geq -r_i$. Fig. 2 shows a constraint graph with 3 nodes representing $x_1$ to $x_3$, where node 0 is the root node. A loop in the constraint graph is nonpositive if the sum of all the edge weights of the loop is nonpositive. Without considering the root node, the constraint graph is actually a register connection graph, where each node represents an FF and an edge represents the combinational connection between FFs, with $w_{ij}$ as the sum of the maximum path delay and setup time and $T$ as the timing budget. The constraint graph is also used in [21] to determine the skew range in static timing analysis. The detailed proof of the equivalence of the nonpositive loop constraint in the graph and the existence of a solution for (1) and (2) can be found in [24].

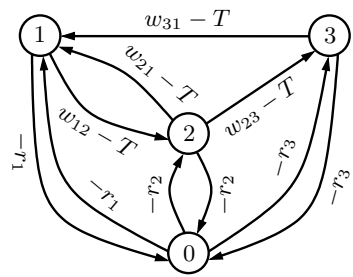

Fig. 2. Constraint Graph

## III. THE PROPOSED METHOD

In this section we will first explain the basic concept to extract $T_{min}$ for the clock period from the constraint graph. Thereafter, the criticality computation in the presence of PST buffers will be shown. Finally, several techniques are explained to reduce the runtime.

### A. Extracting $T_{min}$ using graph transformation

The minimum clock period $T_{min}$ must guarantee that there is at least a solution for the difference constraints of (1) and (2). This is equivalent that all the loops in the constraint graph are nonpositive. For convenience we write all the edge weights in the constraint graph into the form $w_{ij} - k_{ij}T$. For edges which are connected with the root node $k_{ij} = 0$ and $w_{ij} = -r_i$. The weight of a loop $l$ in the graph, denoted by $W_l$, is computed by $W_l = \sum_{i,j}(w_{ij} - k_{ij}T)$, where the sum is computed over all the edges in the loop. The nonpositive constraint for the loop $l$ specifies that

$$W_l = \sum_{i,j}(w_{ij} - k_{ij}T) \leq 0 \iff T \geq \sum_{i,j} w_{ij} \Big/ \sum_{i,j} k_{ij} = T_l \quad (3)$$

$T_l$ is called *loop constraint* henceforth because it is a lower bound for the clock period. In the constraint graph the loops containing only edges connected with the root node have $\sum_{i,j} k_{ij} = 0$. These loops are not included in (3) because the corresponding $\sum_{i,j} w_{ij}$ are always negative.

The constraint (3) from one loop creates a lower bound for the feasible clock period. If all loops are considered, the minimum clock period $T_{min}$ can be computed as

$$T_{min} = \max_{l \in L} T_l \quad (4)$$

where $L$ is the set of all loops in the graph except those which only include the edges connected with the root node. These loops always have negative loop weights and create no constraint for $T_{min}$.

To compute $T_{min}$ in (4) the direct enumeration of all loops is prohibitive owing to the number of loops in a large constraint graph. Instead, we use a method based on graph transformation to capture the loops whose weights affect $T_{min}$. These loops tend to have fewer numbers of nodes on them than the theoretically largest loop, which contains all nodes in the graph. This observation will be explained in Section III-C. The basic concept of traversing loops by graph transformation has been used in [25] for statistical timing analysis of level-sensitive circuits. However, it can not handle the constraint graph, especially due to the full connections from and to the root node. In our method several improvements will be introduced to handle the constraint graph and an edge tracing method will be used to compute the criticalities.

Two basic graph merge operations are used in the graph transformation: serial merge and parallel merge. Fig. 3 illustrates an example of the *serial merge* operation. The transformation removes node $v$ from the graph and creates direct edges between its input and output nodes. The weight of a new edge is equal to the sum of the weights of edges from which the new edge is formed and is also in the same form $w_{ij} - k_{ij}T$, so that the serial merge operation can be applied repeatedly. In this transformation, an input node and an output node may be the same. For example, if we apply the serial merge to node 1 in Fig. 2, a new edge will be created forming a loop at node 2 and similarly at the root node. Such a loop, called *self loop*, is removed from the graph and the constraint from it, in the from of (3), is extracted and merged into $T_{min}$ using (4). Because the serial merge operation creates direct edges between nodes repeatedly, self loops actually contain the sum of the edge weights from the original graph. For example after removing node 1, the self loop at node 2 captures the constraints from the loop $2 \to 1 \to 2$ in Fig. 2. If there are other loops which pass through the removed node $v$, serial merge does not affect their loop weights because the edge weights from any input to any output of $v$ are maintained in the new edges, so that serial merge can be applied repeatedly until all loop constraints are extracted and $T_{min}$ is computed eventually.

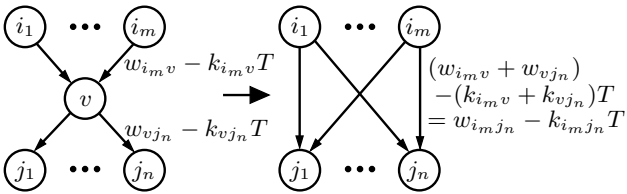

Fig. 3. Serial Merge

After each serial merge operation the node number of the graph is reduced by one. If a node with $m$ inputs and $n$ outputs is removed, $m \times n$ new edges are created, assuming no self loops appear. Normally this is far larger than the number of the removed $m + n$ edges, thus causing the edge number in the graph to increase very fast during the transformation. Actually applying the serial merge operation repeatedly to capture all loop constraints enumerates all the loops, which is prohibitive for a large graph. Therefore, further methods will be applied to reduce the number of edges after each serial merge operation.

Because new edges between the input and output nodes are created during the serial merge operation, parallel edges may appear, which can be merged together to reduce the number of edges. For example, if node 3 in Fig. 2 is removed, a new edge is created from node 2 to 1, parallel to the edge from 2 to 1 in the original graph. An example of the parallel edges is shown in the left of Fig. 4. If the coefficients of $T$, $-k_1$ and $-k_2$ in Fig. 4, in the weights of the parallel edges are equal, these edges can be merged into one edge by the operation called *parallel merge*. In Fig. 4, the two edges with $-k_1T$ are merged, but the other edge with $-k_2T$ can not be processed similarly because at this time we do not know the value of $T$. Similar to the serial merge operation, parallel merge does not affect the constraints from the weights of loops which pass through $i$ and $j$, because the maximum weight of the loops through the merged edges is maintained by the new edge.

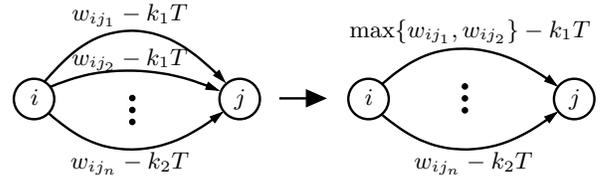

Fig. 4. Parallel Merge

Using the two graph transformation operations we can capture all loop constraints and compute the minimum clock period $T_{min}$, shown in Algorithm 1. The function $remove\_self\_loop(i)$ removes all edges from and to $i$, which are self loops generated after the transformation. These loops are formed by compressing the loops comprising multiple edges in the original graph by repeatedly applying the serial merge operation. For each of these self loops the constraint in (3) is computed and $T_{min}$ is updated using (4), therefore capturing the constraints of the loops in the original graph. Although the parallel merge operation can reduce the edge number

---

**Algorithm 1**: Computing $T_{min}$ using Graph Transformation

**L1**    **while** *more than one node exists in the graph* **do**
**L2**        next_node=select_node();
**L3**        serial_merge(next_node);
**L4**        **foreach** *node $i$ in the input nodes of next_node* **do**
**L5**            remove_self_loop($i$);
**L6**        **end**
**L7**        **if** *parallel edges exist between nodes $i$ and $j$* **then**
**L8**            parallel_merge($i$, $j$);
**L9**            compress_parallel_edges($i$, $j$);
**L10**      **end**
**L11** **end**

after each serial transformation, the edge number in the graph still increases very fast. In Algorithm 1 the functions $select\_node()$ and $compress\_parallel\_edges()$ use some heuristics to reduce the edge number and the runtime. These techniques will be explained in Section III-C.

## B. Computing criticality in the presence of PST buffers

For circuit optimization the timing analysis tool should report a set of gates which are critical to the circuit performance. The probability that a gate delay affects the circuit performance is called *criticality* [3], [17], [19]. Because PST buffers allow the path delays to compensate each other across FFs, the critical paths may span more than one stage of FFs. Fig. 5 shows an example of such critical paths, where the inverters represent the combinational delays shown above the gate. If PST buffers are not considered, the critical path is between FFs 2 and 3. However, the paths between FFs 1 and 3 allow a minimum clock period of 5 due to PST buffers having a range of up to 3. In contrast the minimum clock period constrained by the paths between FFs 4 and 6 is 6, which form the critical paths in this circuit. To capture the critical paths with PST buffers we first define the criticality for loops and edges in the constraint graph. Thereafter, the concept in [17] is extended to include this information into the computation of criticality for combinational gates.

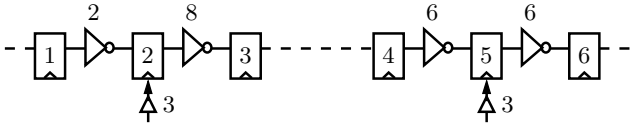

Fig. 5. Critical Paths with PST Buffers

According to (4) the circuit performance is constrained by all the loop constraints $T_l$. Because edge weights are random variables, any loop has a probability to dominate the circuit performance, defined as

$$c_l = prob\{T_l \geq T_{min}\} \quad (5)$$

The larger $c_l$ is the more effect the loop has on the circuit performance. Therefore $c_l$ is called *loop criticality*. A loop with a large $c_l$ is called *critical loop* and the edges on a critical loop are candidates for optimization.

In the constraint graph an edge may be on multiple loops. If any of these loops dominate the circuit performance, the edge is critical. Therefore the criticality for an edge $e$, called *sequential criticality*, is defined as

$$c_e = prob\{\bigvee_{l \in L_e} (T_l \geq T_{min})\} \quad (6)$$

$$= prob\{\neg(\bigwedge_{l \in L_e} (T_l < T_{min}))\} \quad (7)$$

$$= prob\{\neg(\max_{l \in L_e}\{T_l\} < T_{min})\} \quad (8)$$

$$= prob\{\max_{l \in L_e}\{T_l\} \geq T_{min}\} \quad (9)$$

where $L_e$ is the set of loops across $e$ and $\max_{l \in L_e}\{T_l\}$ is the maximum constraint of all these loops. $\wedge$ means *logic and*, $\vee$ *logic or* and $\neg$ *logic not*.

An edge in the constraint graph, if not connected to the root node, corresponds to the maximum delay between a pair of FFs in the circuit. Because the gates between a pair of FFs form many combinational paths, not all gate delays are critical even if the sequential criticality $c_e$ is large. In [17], [18] the cutset concept is introduced to define the criticality of a gate delay. This concept is illustrated in Fig. 6. Considering the combinational circuit between FFs $i$ and $j$, all the paths can be partitioned into two sets: $P_g$ and $P_{\bar{g}}$. $P_g$ contains the paths between $i$ and $j$ and across the gate $g$; $P_{\bar{g}}$ contains all the other paths between $i$ and $j$. The gate delay is critical if the longest path passes across it, that is, it belongs to $P_g$. The maximum path delay $d_{P_g}$ from $P_g$ can be computed by $d_{ig} + d_g + d_{gj}$, where $d_{ig}$ is the maximum delay from the output $i$ to the input of $g$, and $d_{gj}$ the delay from the output of $g$ to the input of $j$. $d_g$ is the gate delay. According to [18], the criticality of the gate delay is computed by

$$c_g^{no\_pst} = prob\{d_{P_g} \geq d_{P_{\bar{g}}}\} \quad (10)$$

$$= prob\{d_{P_g} \geq \max\{d_{P_g}, d_{P_{\bar{g}}}\}\} \quad (11)$$

$$= prob\{d_{P_g} \geq d_{ij}\} \quad (12)$$

where $d_{P_{\bar{g}}}$ is the maximum path delay from $P_{\bar{g}}$; $d_{ij}$ is the maximum path delay between $i$ and $j$. Detailed proof of this computation can be found in [17], [18].

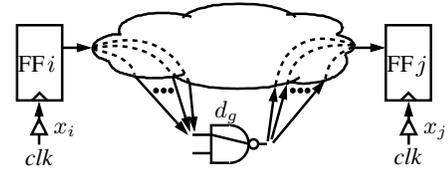

Fig. 6. Criticality of Combinational Delays

The criticality $c_g^{no\_pst}$ defines the probability that the paths across a gate dominate the other paths. When PST buffers are considered, this gate is critical only if the delay between $i$ and $j$, which corresponds to an edge in the constraint graph, is on a loop $l$ which dominates $T_{min}$, that is, $T_l \geq T_{min}$. A gate $g$ may be on the combinational paths between many pairs of FFs, corresponding to a set of edges, written as $E_g$, in the constraint graph. Combining (6)-(9) and (12), the *criticality of a gate delay* in the presence of PST buffers is defined as

$$c_g = prob\{\bigvee_{e \in E_g}(\max_{l \in L_e}\{T_l\} \geq T_{min} \wedge d_{P_g} \geq d_{ij})\} \quad (13)$$

$$= prob\{\bigvee_{e \in E_g}(0 \geq \max\{T_{min} - \max_{l \in L_e}\{T_l\}, d_{ij} - d_{P_g}\})\} \quad (14)$$

where edge $e$ is between nodes $i$ and $j$ in the constraint graph; $L_e$ is the set of loops passing through edge $e$ in the graph; $d_{ij}$ is the maximum path delay between FFs corresponding to $i$ and $j$. Let $C_e = \max\{T_{min} - \max_{l \in L_e}\{T_l\}, d_{ij} - d_{P_g}\}$,

$$c_g = prob\{\neg(\bigwedge_{e \in E_g}(C_e > 0))\} \quad (15)$$

$$= prob\{\neg(\min_{e \in E_g}\{C_e\} > 0)\} \quad (16)$$

$$= prob\{\min_{e \in E_g}\{C_e\} \leq 0\} \quad (17)$$

In practice only the edges with sequential criticality $c_e$ defined in (9) larger than 0 in the edge set $E_g$ should be processed for the computation from (13) to (17) so that the runtime can be reduced.

The criticality $c_g$ measures the probability that a combinational gate affects the circuit performance $T_{min}$. However, in the process space only some of the chips may fail in respect to a given clock period $T_s$ from the timing specification of the design. It is therefore reasonable to optimize the critical paths and gates in these chips to meet the timing specification. The criticality of a gate delay against a given clock period $T_s$ is defined in [17] to compute $c_g^{no\_pst}$ in

(10) under the condition $T_{min} > T_s$. Similarly the criticality defined in (13)-(17) can be extended to incorporate this condition. This conditional criticality, denoted by $c_g^{T_s}$, is computed as follows.

$$c_g^{T_s} = prob\{\min_{e \in E_g}\{C_e\} \leq 0 \mid T_{min} > T_s\} \quad (18)$$

$$= prob\{\min_{e \in E_g}\{C_e\} \leq 0 \ \wedge \ T_{min} > T_s\}/prob\{T_{min} > T_s\} \quad (19)$$

$$\cong prob\{\max\{\min_{e \in E_g}\{C_e\}, T_s - T_{min}\} < 0\}/prob\{T_{min} > T_s\} \quad (20)$$

where $T_s$ is the given clock period. (19) and (20) are roughly equal owing to the fact that the existing statistical timing engines approximate the $max$ and $min$ computations using continuous functions.

To compute the criticalities several variables should be known in (6)-(9) and (13)-(20). The minimum clock period $T_{min}$ can be computed using Algorithm 1. The path delays $d_{ij}$ and $d_{P_g}$ can be computed using an SSTA engine as described in [17]. The computation of $\max_{l \in L_e}\{T_l\}$ for an edge $e$ in the constraint graph however needs to traverse the loops passing through the edge $e$, which are normally in a prohibitive number.

Instead of searching the loops for each edge individually, we apply the serial merge operation to the original constraint graph again after $T_{min}$ is computed. For each edge created during the transformation, we maintain an *edge tracing list* to trace the edges which form the new edge. When two consecutive edges are replaced by a new edge in the serial merge operation in Fig. 3, the edge lists in the replaced edges are copied into the edge list of the new edge, denoted by the function *copy_edge_lists()*. Each time when a self loop is formed, the loop is removed and the loop constraint $T_l$ defined in (3) is updated into the variable holding the value $\max_{l \in L_e}\{T_l\}$ for each edge in the edge tracing list. The computation of $\max_{l \in L_e}\{T_l\}$ is summarized as follows.

| **Algorithm 2**: Computing $\max_{l \in L_e}\{T_l\}$ for Edges |
|---|
| L1  **while** *more than one node exists in the graph* **do** |
| L2     next_node=select_node(); |
| L3     serial_merge(next_node); |
| L4     **foreach** *newly created edge* **do** |
| L5        copy_edge_lists(); |
| L6     **end** |
| L7     **foreach** *node i in the input nodes of next_node* **do** |
| L8        remove_self_loop(i); |
| L9        **foreach** *removed edge e forming a self loop* **do** |
| L10          **foreach** *edge in the edge tracing list of e* **do** |
| L11             update_loop_constraint(); |
| L12          **end** |
| L13       **end** |
| L14    **end** |
| L15    **if** *parallel edges exist between nodes i and j* **then** |
| L16       compress_parallel_edges($i$, $j$); |
| L17    **end** |
| L18    **if** *exit_condition()* **then** |
| L19       exit; |
| L20    **end** |
| L21 **end** |

The main structure of Algorithm 2 is similar to Algorithm 1 because they both capture the nonpositive loop constraints. The difference is that Algorithm 1 updates the constraints into $T_{min}$ using (4), but Algorithm 2 updates the constraints into the variables holding $\max_{l \in L_e}\{T_l\}$ for edges which are parts of the loop in the original graph. This update operation is denoted by the function *update_loop_constraint()* in Algorithm 2. Because the new edges have different edge tracing lists, the parallel merge operation in Fig. 4 can not be applied even if they have the same coefficients of $T$. This limitation increases the edge number during the transformation and causes the runtime larger than Algorithm 1. In the next section, we will explain several techniques to reduce the runtime of both algorithms, including the functions *compress_parallel_edges()*, *select_node()* and *exit_condition()* in Algorithm 1 and 2.

## C. Implementation issues

In the serial merge operation in Fig. 3 the number of edges increases in most cases. This not only causes the runtime in the following transformations to increase but also consumes much memory. In Algorithm 2 the parallel merge operation can not be applied because even if the edge weights have equal coefficients of $T$ they may have different edges in their edge tracing lists. To solve this problem, we compare the weights of parallel edges. If an edge has a much smaller delay than the other, it is removed from the constraint graph. Because all edge weights are random variables, the comparison can only be performed statistically. Consider two edge weights $w_1 - k_1 T$ and $w_2 - k_2 T$, the second edge will be dominated if

$$prob\{w_1 - k_1 T \geq w_2 - k_2 T\} > \sigma_h \quad (21)$$

where $\sigma_h$ is a predefined value near to 1. In Algorithm 2 the clock period is known from Algorithm 1, so that the probability can be computed directly. However, in Algorithm 1 the constraints created by the loops are updated into $T_{min}$ using (4). Before the algorithm is finished, the value of $T_{min}$ is only temporary and should be smaller than the final value because there are loop constraints which have not been processed. Therefore, we compare the parallel edges, if $k_2 - k_1 > 0$, by computing

$$prob\{T_{min} \geq (w_2 - w_1)/(k_2 - k_1)\} \quad (22)$$

If the probability in (22) is larger than $\sigma_h$, edge weight $w_2 - k_2 T$ is dominated by $w_1 - k_1 T$. This edge removal technique is denoted by *compress_parallel_edges()* in Algorithm 1 and 2. During this edge compression, special cases, for example the probability masking described in [20], should be handled. Because the circuit performance $T_{min}$ is dominated by the longer paths between FFs, many other paths have relative small delays. Therefore the new edges which are formed from a large number of consecutive edges in the original constraint graph by applying serial merge operations are very likely to be dominated by other parallel edges. This can also be explained by the fact that in most cases a long path delay in the circuit needs not to be compensated by the other FF stages far from it.

The second issue is the order of nodes for the serial merge operation. If a node with $m$ inputs and $n$ outputs is removed by the serial merge operation, $m \times n$ edges are created. An extreme case is the root node, which has edges to and from all other nodes. If the root node is removed using the serial merge operation, for any two nodes in the graph two new edges are created. In a graph with many nodes this will generate a large number of new edges, which are impossible to be processed by the following transformations. Although an optimal node processing order, which guarantees the number of edges in each of the transformations is minimal, may exist, to find this optimal order is very difficult and, even if possible, consumes much runtime. In the proposed method, we use a heuristic algorithm to select the next node for the serial merge operation. First we select the nodes with the smallest *node connection*, which is defined as $n_i \times n_o$ for a node, where $n_i$ and $n_o$ are numbers of nodes which have edges to and from the node under evaluation respectively. If there are multiple nodes with the same smallest node connection, the node with the fewest edges is selected. This node selection is heuristic but very effective in the graph transformation. Because the root node has edges to and from all other nodes, it is excluded from the node selection until it is the only node in the graph. At this time no edge exists in the graph because all edges have been processed and removed. The node selection is denoted by the function $select\_node()$ in Algorithm 1 and 2.

The third issue is the exit condition, the function $exit\_condition()$ in Algorithm 2. In the parallel removal technique above, even though some edges are already negative, they may not be dominated by each other. To handle this case, all the edges in the graph are checked if they are negative by the function $exit\_condition()$ after each serial merge operation. If all edges are negative, it is guaranteed that no positive loop is left in the graph. Because the edge weight is a random variable, the negativity of an edge in the constraint graph is defined by

$$prob\{w_{ij} - k_{ij}T < 0\} > \sigma_h \quad (23)$$

The last issue is the fact that not every FF but the ones which are relevant to critical paths have PST buffers. If an FF does not have a PST buffer, the corresponding variable $x_i$ in (1) is 0. If a constraint in (1) contains only one variable, an edge is created from or to the root node, similar to the case for (2). If both variables in (1) are 0, the constraint simply creates a lower bound for the clock period and no edge needs to be created in the constraint graph. A special case is that a path from and to the same FF. In this case the two variables in (1) are the same, so that the constraint also creates a simple lower bound for the clock period. According to this observation, many nodes corresponding to FFs need not to appear in the constraint graph. This not only reduces the numbers of the nodes and edges but also breaks large loops, making the timing analysis and criticality computation faster.

## IV. EXPERIMENTAL RESULTS

The algorithms were implemented in C++ and tested using a 2.33GHz CPU. The ISCAS89 benchmark circuits were used for experiments, where all FFs were assigned PST buffers. The gates in the benchmark circuits were mapped to a 90nm library from an industry partner. The standard deviations of transistor length, oxide thickness and threshold voltage were assigned to 15.7%, 5.3% and 4.4% of the nominal values respectively [26]. The gate delays were created using the method proposed in [1]. We used the SSTA engine proposed in [3] to compute the sum and maximum of random variables. In the proposed method the probability threshold $\sigma_h$ in (21) and (23) was set to 0.99 to guarantee accuracy.

To verify the accuracy of the proposed method, we ran Monte Carlo simulation with 10 000 samples. For each sample, the minimum clock period constrained by (1) and (2) was computed using linear programming. The distribution formed by all the performance samples was compared with $T_{min}$ computed by the proposed method and the results are shown in Table I, with the PST ranges set to 1/8 of the clock period calculated without considering PST buffers, roughly the range used in [11].

In Table I $n_c$ and $n_s$ denote the numbers of combinational cells and sequential cells in the circuit respectively. Next, the accuracy of the proposed method compared to Monte Carlo simulation is shown. $\mu_{err}$ is the relative error of the mean of the minimum clock period $T_{min}$ and defined as $|\mu_{SSTA} - \mu_{MC}|/\mu_{MC}$, where $\mu_{SSTA}$ and $\mu_{MC}$ are the means of the minimum clock period computed by the proposed method and by Monte Carlo simulation using the original circuit respectively. Similar to $\mu_{err}$, $\sigma_{err}$ is defined to show the accuracy of the standard deviation of the clock period. From $\mu_{err}$ and $\sigma_{err}$ we can see that the results of the proposed method are accurate for statistical timing analysis.

TABLE I
RESULTS OF SSTA FOR CIRCUITS WITH PST BUFFERS

| Circuit | $n_c$ | $n_s$ | $\mu_{err}$ | $\sigma_{err}$ | $t_P$ (s) | $t_{MC}$ (s) | Speedup |
|---|---|---|---|---|---|---|---|
| s298 | 119 | 14 | 0.18% | 0.64% | 0 | 8.95 | |
| s526 | 193 | 21 | 0.20% | 0.36% | 0 | 15.2 | |
| s820 | 289 | 5 | 0.22% | 0.17% | 0 | 11.59 | |
| s1238 | 508 | 18 | 0.23% | 0.28% | 0 | 10.59 | |
| s1423 | 657 | 74 | 0.27% | 0.76% | 0.31 | 465.72 | 1502 |
| s5378 | 2779 | 179 | 0.23% | 0.32% | 0.05 | 1608.87 | 32177 |
| s9234 | 5597 | 211 | 0.56% | 0.44% | 0.11 | 1681.25 | 15284 |
| s13207 | 7951 | 638 | 0.41% | 0.42% | 0.10 | 2107.82 | 21078 |
| s15850 | 9772 | 534 | 1.00% | 0.39% | 3.66 | 9690.81 | 2648 |
| s38584 | 19253 | 1246 | 1.26% | 0.65% | 7.56 | 21705.9 | 2871 |

The runtimes of timing analysis using the proposed method and Monte Carlo simulation, which is used in the existing methods for timing analysis and optimization of circuits with PST buffers, are shown as $t_P$ and $t_{MC}$ in seconds respectively. The 0s in Table I represent that the runtimes are shorter than $10^{-6}$s and can not be measured accurately using the *clock* function. The speedup of the proposed method compared to Monte Carlo simulation is shown in the Speedup column of Table I. Since Monte Carlo simulation is the only method available for statistical timing analysis of circuits with PST buffers, this comparison demonstrates the efficiency of the proposed method and its applicability to accelerate other methods which depend on the results of statistical timing analysis.

Using PST buffers the circuit performance can be improved, as explained in Section II. However, the performance improvement is bounded because after the range of PST buffers reaches a threshold the circuit performance is determined by the maximum average edge delay between FFs across loops in the circuit, which does not change as the PST range changes, and therefore can not be improved by further cycle stealing. To show the relation of the circuit performance and the PST range, we tested the circuit performance by setting the size of the PST range $(-r_i, r_i)$ in (2), equal to $2 \times r_i$, to 1/32, 1/16, 1/8, 1/4, 2/5 and 2/3 times of the clock period calculated without considering PST buffers, denoted by $T_{no\_pst}$. The trend of the mean values of the minimum clock period is shown in Fig. 7. As the range of the PST buffers increases, the clock period decreases, meaning higher performance, because the delays between FFs are allowed to compensate each other in a larger range. According to Fig. 7, PST buffers can reduce the clock period by about 30%, and therefore have great value in high performance designs. However, when the PST range reaches about 40% of the clock period calculated without PST buffers, no significant further circuit performance can be gained. Owing to the efficiency of the proposed method, it can be used to evaluate the relation between the minimum clock period and the range of PST buffers, allowing designers to make tradeoff between performance and die size.

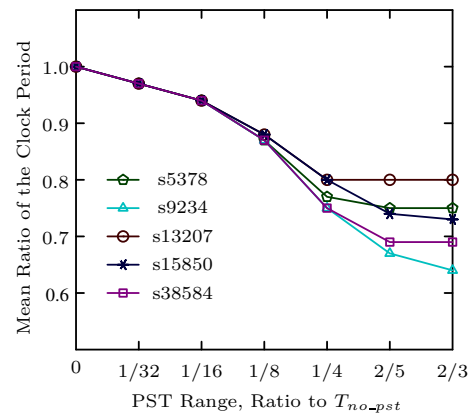

Fig. 7. Trend of the Mean Values of the Minimum Clock Period

To verify the proposed criticality computation, the distances between nodes in the constraint graph were computed using the Bellman-Ford algorithm [24] in each Monte Carlo sample. Thereafter each edge was checked if the loop across it determines the minimum clock period computed by linear programming. Table II shows the results of the comparison. In this test the PST range was set to 1/8 of the clock period calculated without PST buffers. The runtime of the Monte Carlo simulation for the circuit s38584, longer than nine days, is estimated from ten iterations. Because of the large runtime of Monte Carlo simulation, the results of this circuit are not compared, although the proposed method can compute the criticalities in less than 40 seconds. Owing to the approximation in the statistical computations of SSTA engines, the criticality can not be calculated accurately. However, because the purpose to compute criticality is to select the gates for optimization, we compare the sets of critical gates selected by Monte Carlo simulation and the proposed method. In Table II the columns >0.3 and >0.2 show the numbers of gate delays with criticalities larger than 0.3 and 0.2 respectively. $n_c$ is the number of gate delays identified by Monte Carlo simulation, and $n_m$ is the number of gate delays which are missed by the proposed method. The criticality $c_g^{T_s}$ is computed against the specified clock period $T_s$, which is set to $\mu + \sigma$ where $\mu$ and $\sigma$ are the mean and standard deviation of the minimum clock period $T_{min}$ computed by Algorithm 1.

TABLE II
RESULTS OF CRITICALITY COMPUTATION

| Circuit | $c_g, n_m/n_c$ | | $c_g^{T_s}, n_m/n_c$ | | Runtime (s) | |
|---|---|---|---|---|---|---|
| | > 0.3 | > 0.2 | > 0.3 | > 0.2 | $t_P$ | $t_{MC}$ |
| s298 | 0/10 | 0/14 | 0/10 | 0/10 | 0 | 18.84 |
| s526 | 0/10 | 0/10 | 0/10 | 0/10 | 0.01 | 42.51 |
| s820 | 0/12 | 0/12 | 0/12 | 0/12 | 0 | 18.07 |
| s1238 | 0/17 | 0/17 | 0/17 | 0/17 | 0 | 36.51 |
| s1423 | 0/61 | 0/61 | 0/61 | 0/61 | 0.44 | 1466.44 |
| s5378 | 0/31 | 0/45 | 0/31 | 0/38 | 1.09 | 4227.07 |
| s9234 | 0/71 | 2/93 | 0/75 | 2/113 | 1.12 | 15338.9 |
| s13207 | 0/32 | 1/69 | 0/37 | 0/70 | 2.05 | 84551.4 |
| s15850 | 0/32 | 63/133 | 0/36 | 23/89 | 14.51 | 211356 |
| s38584 | - | - | - | - | 39.80 | >9d |

According to Table II, the proposed method can identify the gate delays with criticalities larger than 0.3 effectively. However for the circuit s15850 many critical gate delays are missed when the bound for the selection is lowered to 0.2. This is because many criticalities computed by the proposed method fall through the boundary 0.2. For example, if the sets of gates with criticalities larger than 0.01 are compared, the proposed method can identify 449 gates out of 473 gates which are identified by Monte Carlo simulation. In the missing gates the maximum criticality is 0.02. This shows that for optimization purpose the proposed method can provide enough information, especially in view of the very short runtime. As pointed out in [17], the skewness of the distribution, which is not considered by many SSTA engines, may cause large inaccuracy in criticality computation. Especially when the critical paths are compared to the circuit performance, the criticality is very sensitive to the inaccuracy of the statistical approximations. For example, in (12) the inaccuracy of $d_{P_g}$ and $d_{ij}$ may cause a large deviation of the criticality from the correct value. In order to improve the accuracy of criticality computation, further methods, for example [27], can be considered.

## V. CONCLUSION

In this paper, we propose a fast method to compute the minimum clock period for circuits with PST buffers. Using graph transformation the proposed method can effectively capture the constraints from loops in the constraint graph. The resulting minimum clock period is in a parametric form from which we can compute the yield of the circuit for any given clock period. The criticalities of gate delays are also computed, providing information for circuit optimization.